\documentclass[lettersize,journal]{IEEEtran}
\usepackage{amsmath,amsfonts}
\usepackage{algorithmic}
\usepackage{array}
\usepackage[caption=false,font=footnotesize,labelfont=rm,textfont=rm]{subfig}
\usepackage{textcomp}
\usepackage{stfloats}
\usepackage{url}
\usepackage{verbatim}
\usepackage{graphicx}
\usepackage{amsmath}
\usepackage{amssymb}
\usepackage{bm}
\usepackage{graphicx}
\usepackage{filecontents,lipsum}
\usepackage{paralist}
\usepackage{enumitem}
\usepackage{color}
\usepackage{cite}
\usepackage{geometry}
\usepackage{diagbox}
\usepackage{booktabs}
\usepackage{stfloats}
\usepackage{makecell}
\usepackage{multirow}
\usepackage{algorithm}
\usepackage{algorithmic}
\usepackage{hyperref}
\usepackage{stfloats}
\usepackage{overpic}
\usepackage{subfig}
\usepackage{cleveref}
\usepackage{amsmath}
\usepackage{amsmath, amssymb}

\def\BibTeX{{\rm B\kern-.05em{\sc i\kern-.025em b}\kern-.08em
    T\kern-.1667em\lower.7ex\hbox{E}\kern-.125emX}}
\usepackage{balance}
\begin{document}
\title{Improving Channel Resilience for Task-Oriented Semantic Communications: A Unified Information Bottleneck Approach}
\author{Shuai Lyu, Yao Sun,~\IEEEmembership{Senior Member,~IEEE}, Linke Guo,~\IEEEmembership{Member,~IEEE}, Xiaoyong Yuan, Fang Fang,~\IEEEmembership{Senior Member,~IEEE}, Lan Zhang,~\IEEEmembership{Member,~IEEE}, Xianbin Wang,~\IEEEmembership{Fellow,~IEEE}

\thanks{This work was supported by the US National Science Foundation (NSF) under Grants CCF-2106754, CCF-2221741, CCF-2418308, and CCF-2151238.} 
\thanks{Shuai Lyu, Linke Guo, Xiaoyong Yuan, and Lan Zhang (the corresponding author) are with the Department of Electrical and Computer Engineering, Clemson University, SC, USA (e-mail: \{slv; linkeg; xiaoyon; lan7\}@clemson.edu).}
\thanks{Yao Sun is with the James Watt School of Engineering, University of Glasgow, Glasgow, U.K. (e-mail: Yao.Sun@glasgow.ac.uk).}
\thanks{Xianbin Wang and Fang Fang are with the Department of Electrical and Computer Engineering, and Fang Fang is also with the Department of Computer Science, Western University, London, ON, Canada (e-mail: \{xianbin.wang; fang.fang\}@uwo.ca).
}}

\markboth{Under review  for IEEE COMMUNICATIONS LETTERS}%
{Shell \MakeLowercase{\textit{et al.}}: Bare Demo of IEEEtran.cls
for Journals}

\maketitle

\begin{abstract}
Task-oriented semantic communications (TSC) enhance radio resource efficiency by transmitting task-relevant semantic information. However, current research often overlooks the inherent semantic distinctions among encoded features. 
Due to unavoidable channel variations from time and frequency-selective fading, semantically sensitive feature units could be more susceptible to erroneous inference if corrupted by dynamic channels. Therefore, this letter introduces a unified channel-resilient TSC framework via information bottleneck. This framework complements existing TSC approaches by controlling information flow to capture fine-grained feature-level semantic robustness. Experiments on a case study for real-time subchannel allocation validate the framework's effectiveness.
\end{abstract}


\IEEEpeerreviewmaketitle

\section{Introduction}
\IEEEPARstart{F}{uture} wireless networks are expected to support dramatically increased data traffic, driven primarily by the prevalence of sensing capabilities, distributed computing resources, and ongoing convergence with vertical applications, including transportation, online gaming and smart utilities.  To support the unprecedented traffic growth with limited radio resources, task-oriented semantic communications (TSC) have garnered considerable interest~\cite{getu2024survey,gunduz2022beyond}. Unlike conventional bit-level communications, TSC leverages both distributed computing resources as well as the most recent artificial intelligence (AI) techniques to extract and transmit task-relevant semantic information, thereby reducing unnecessary traffic and enhancing radio resource utilization efficiency.

In TSC, a transceiver typically integrates semantic and channel coding functions to extract task-relevant semantic features and enable efficient transmissions~\cite{gunduz2022beyond}. Yang et al. introduced a novel compression method for AI tasks to improve transmission efficiency, demonstrated with a prototype for surface defect detection~\cite{com5}. Instead of focusing solely on semantic coding, Shao et al.~\cite{com4} and Sun et al.~\cite{com10} proposed to jointly optimize semantic and channel coding based on information bottleneck to balance semantic distortion and transmission efficiency. Moreover, TSC problems have been investigated in multi-user and multi-modality scenarios~\cite{xie2021task,liu2023adaptable,com16}.

Despite considerable efforts, existing TSC research primarily focuses on statistical channel conditions, assuming a certain channel condition when transmitting an input sample~\cite{com4,com5,com10,xie2021task,liu2023adaptable,gunduz2022beyond}.
However, the encoded features may encounter distinct physical impairments during transmission. For instance,  in orthogonal frequency division multiplexing (OFDM) systems, multiple subcarriers with different frequency-selective fading are used for signal transmission across a wide bandwidth~\cite{coleri2002channel}. Consequently, if semantically sensitive feature units are allocated to poorly performing subcarriers, the corrupted features are more susceptible to erroneous task inference. While channel estimation techniques can evaluate instantaneous channel conditions in practice~\cite{coleri2002channel}, the evaluation results cannot effectively guide feature-level transmissions due to the neglected semantic distinctions among encoded feature units. 

To bridge the gap between feature-level semantic distinctions and channel variations, this letter introduces an innovative TSC framework to improve channel resilience by evaluating and prioritizing encoded feature units of input data based on their robustness against channel variations. This framework is designed to be complementarily leveraged by existing TSC approaches to capture fine-grained feature-level semantic robustness, thereby adjusting transmission strategies for channel-resilient TSC. The primary contributions of this letter are summarized below.
\begin{itemize}
\item \emph{Unified channel-resilience framework:} We develop a unified approach to analyze a well-trained TSC transceiver for channel resilience, providing a soft robustness mask for the encoded feature space without modifying the established TSC encoding and decoding functions. This mask will be utilized to prioritize robust feature units and adapt the transmission strategies against instantaneous channel variations in practice. 

\item \emph{Robustness mask based on information bottleneck (IB):} We construct the robustness mask for encoded feature units by leveraging IB to regulate information flow with explicitly added artificial noise. Based on the task inference sensitivity, this mask softly disentangles the encoded features into robust and non-robust from the semantic level. 

\item \emph{Numerical evaluation.} We conduct experiments for real-time subchannel allocation problems as a case study. Evaluation results under two image tasks demonstrate the framework's effectiveness, especially under highly dynamic adverse channel conditions.
\end{itemize}

\section{System Model}
This paper considers a typical task-oriented semantic communications (TSC) system. As illustrated in Fig.~1, the source transceiver comprises semantic and channel encoders. Given an input sample $x\in\mathcal{X}_T$ of task $\mathcal{T}$ with the inherent task-specific semantic information $y\in\mathcal{Y}_T$, \textit{e.g.}, the target label, the source transceiver first extracts the semantic information of $x$ with the semantic encoder and then processes it via the channel encoder. The encoded features are given by $z = E_\varphi(x)$, which is also represented as $z=\{z_1, \dots, z_m\}$, consisting $m$ vectors. Note that we denote $E_\varphi$ as the joint semantic and channel encoding function for ease of representation, which is consistent with many existing TSC works~\cite{com4,ma2023task,com16}. Encoded features $z$ are then transmitted through the physical channel to the destination transceiver, and the received signal can be given by $\hat{z} = Hz+n$, $\hat{z}=\{\hat{z}_1, \dots, \hat{z}_m\}$, where $H$ denotes the channel matrix and $n\sim \mathcal{N}(0,\gamma^2 I)$ denotes the additive white Gaussian noise (AWGN). The received signal is subsequently processed via a channel decoder and a semantic decoder. Similar to the source transceiver, we denote $D_\theta$ as the joint channel and semantic decoding function and derive the reconstructed semantic information $\hat{y}=D_\theta(\hat{z})$.

The workflow of the above TSC system can be formulated as a probabilistic graphical model: $Y \leftrightarrow X \leftrightarrow Z \leftrightarrow \hat{Z}\leftrightarrow \hat{Y}$. In the following, we use upper-case letters, \textit{e.g.}, $X$, and lower-case letters, \textit{e.g.}, $x$, to represent random matrices and their realizations, respectively. Existing TSC research primarily focused on reducing the size of encoded features $z=\{z_1, \dots, z_m\}$ while ensuring reconstruction performance of semantic information~\cite{com4,com5,com10,xie2021task,liu2023adaptable,gunduz2022beyond}. 
To achieve this, encoding and decoding functions, \textit{i.e.}, $E_\varphi$ at the source transceiver and $D_\theta$ at the destination transceiver, are strategically optimized. We refer readers to recent TSC works~\cite{getu2024survey,com4,com5,com10,xie2021task,liu2023adaptable,gunduz2022beyond} for more details.

\section{Channel-Resilient TSC Framework}
\subsection{Design Intuition}
While TSC systems have been extensively studied~\cite{getu2024survey,com4,com5,com10,xie2021task,liu2023adaptable,gunduz2022beyond}, existing research mainly considers statistical channel conditions to optimize semantic and channel coding functions, assuming that all encoded features of an input sample are transmitted under the same channel condition. However, this assumption may {not hold true} in practice. For instance, in OFDM systems, multiple subcarriers are used to simultaneously transmit data across a wide band~\cite{coleri2002channel}. Hence, \textit{encoded features may suffer distinct channel impairments} due to different frequency-selective fading between subcarriers. Although instantaneous channel state information (CSI) can be monitored using estimation techniques, such as pilot symbols embedded in OFDM symbols, the adaptation of transmission strategies is designed to maximize the successful delivery of OFDM symbols. {The focus remains on optimizing bit-level transmission performance, \textit{i.e.}, duplicating all encoded features, rather than semantic inference}. In other words, \textit{feature units across the encoded signal are assumed equally robust against channel variations}. However, feature units may affect semantic inference differently, \textit{e.g.}, the semantically sensitive or non-robust units are more susceptible to erroneous task inference if corrupted by poor channel conditions. 
To address these limitations, this letter introduces an innovative framework for channel-resilient TSC by evaluating and prioritizing encoded feature units based on their robustness against channel variations. The framework offers a \textit{unified} solution to complement existing TSC approaches from two perspectives: 
\begin{itemize}
    \item The framework seamlessly integrates with existing TSC approaches by analyzing a well-trained TSC transceiver. Specifically, a soft robustness mask is created for the encoded feature space without modifying the established TSC encoding and decoding functions, \textit{i.e.}, $E\varphi$ and $D_\theta$.
    \item The robustness mask aims to align semantic-level feature units with instantaneous channel variations. The mask with feature-level semantic distinctions guides transceivers to efficiently adjust transmission strategies based on instantaneous CSI, aiming for task-specific semantic inference. 
    
\end{itemize}

To achieve this goal, we leverage information bottleneck (IB)~\cite{com2} to analyze the encoded feature space by explicitly adding artificial noise to synthesize channel variations. Based on the semantic inference sensitivity of artificial noise intervention, a soft feature robustness mask is generated to \textit{indicate how encoded feature units corrupted by channel variations affect semantic inference with assigned information}. In the following, we first present the IB reformulation to control information flow for channel resilience purposes and then provide a tractable solution to obtain the robustness mask. Finally, a case study of leveraging the mask for subchannel allocation is introduced. 


\begin{figure}[t!]
\begin{center}
\includegraphics[width=.48\textwidth]{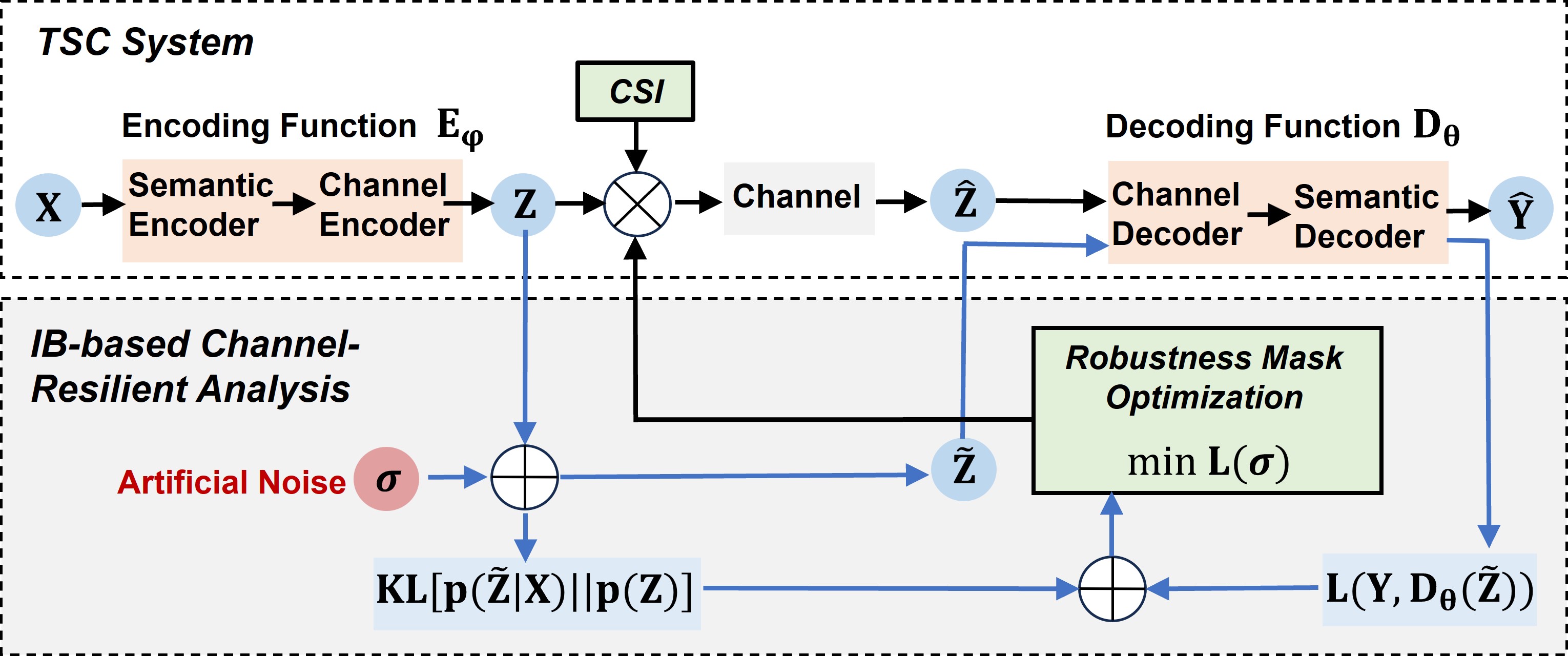}
\vspace{0.1em}
\caption{Overview of channel-resilient TSC framework. Blue arrows indicate IB-based channel-resilient analysis; black arrows indicate transmission procedures in the TSC system.}
\label{fig:figure1}
\end{center}
\vspace{-1.5em}
\end{figure}

\subsection{Information Bottleneck Reformulation}
IB is an information theoretical design principle that aims to find the best tradeoff between accuracy and complexity~\cite{com2}. Several recent works leveraged the IB principle in TSC to seek the tradeoff between transmission rate and semantic information distortion~\cite{com4,com9}. Following the aforementioned probabilistic graphical model in Section II, $Y \leftrightarrow X \leftrightarrow Z \leftrightarrow \hat{Z}\leftrightarrow \hat{Y}$, the objective of optimizing the encoding and decoding functions, \textit{i.e.}, $\varphi$ and $\theta$, based on the IB principle can be given by 
\begin{align}\label{eq:opt1}
\begin{split}
\min L(\varphi,\theta)= - I(\hat{Z},Y) + \beta I(\hat{Z},X),
\end{split}
\end{align}
where 
$I$ is the mutual information; $\beta$ is the Lagrange multiplier that regulates the amount of information in feature space $\hat{Z}$. In (\ref{eq:opt1}), the first term minimizes semantic distortion by allowing the received encoded features $\hat{Z}$ to be predictive on semantic information $Y$, and the second term maximizes transmission rate by enforcing the compression from the input $X$ to received encoded features $\hat{Z}$. 

Instead, this paper intends to improve the channel resilience of a well-trained TSC transceiver, whose encoding and decoding functions $(\varphi,\theta)$ are frozen. To estimate how each encoded feature corrupted by channel variations affect semantic inference, we resue the objective in (\ref{eq:opt1}) by introducing the artificial noise $\sigma$ to control the information flow and estimate the robustness of the encoded features. Thus, the objective function of robustness estimation for channel-resilient TSC can be formulated as
\begin{align}\label{eq:opt2}
\begin{split}
\min & \ \ L(\sigma)= - I(\Tilde{Z},Y|\varphi, \theta) + \beta I(\Tilde{Z},X|\varphi, \theta),
\\s.t. & \ \ \Tilde{Z}=Z+\sigma \cdot \epsilon,
\end{split}
\end{align}
where $\Tilde{Z}$ stands for artificially corrupted encoded features by injecting the artificial noise $\sigma$ to encoded features $Z = E_\varphi(X)$. Hence, we derive $\Tilde{Z}\sim \mathcal{N}(E_\varphi(X),\sigma^2)$. The operator $\cdot$ in (\ref{eq:opt2}) denotes the Hadamard product, and $\epsilon$ represents the Gaussian noise sampled from $\mathcal{N}(0,I)$. The noise variation here measures the correlation between the encoded features and the inferred semantic information based on the fact that robustness refers to a high correlation on semantic inference and non-robustness is opposite~\cite{tsipras2018robustness}. As $(\varphi,\theta)$ are frozen throughout the optimization for channel resilience analysis, we exclude $\varphi$ and $\theta$ in the following notation for simplicity.

\subsection{Upper Bound of the IB-based Robustness Estimation}
To calculate the achievable artificial noise, we resolve the difficulty of mutual information computation in (\ref{eq:opt2}) by deriving the upper bound of the objective function. We start with the first term $I(\Tilde{Z}, Y)$. Writing it out in full, this becomes
\begin{align}
\begin{split}
I(\Tilde{Z},Y)&=\int p(y,\Tilde{z})\log\frac{p(y,\Tilde{z})}{p(y)p(\Tilde{z})}dyd\Tilde{z}\\
&=\int p(y,\Tilde{z})\log p(y|\Tilde{z})dyd\Tilde{z}+H(Y),
\end{split}
\end{align}
where $p(y)$ and $p(\Tilde{z})$ are the probability of the semantic information and encoded features with artificial noise, respectively. $H(Y)$ is the entropy of the semantic information that is independent to the optimization and thus ignored. Recalling the probabilistic graphical model $Y\leftrightarrow X\leftrightarrow Z$, we rewrite $p(y,\Tilde{z})$ based on the underlying characteristics of the Markov chain,
\begin{align}
\begin{split}
p(y,\Tilde{z})=\int p(x)p(z|x)p(\Tilde{z}|z)p(y|\Tilde{z})dxdz.
\end{split}
\end{align}

Then, we derive the new form of $I(\Tilde{Z}, Y)$ as 
\begin{align}\label{eq:IYX_upper}
\begin{split}
I(\Tilde{Z},Y)&\cong\int p(y,\Tilde{z})\log p(y|\Tilde{z})dyd\Tilde{z}\\
&=\underset{X\sim p(X),Z\sim p_\varphi(Z|X), \Tilde{Z}\sim p(\Tilde{Z}|Z)}{\mathbb{E}}[-\mathcal{L}(Y,D_\theta(\Tilde{Z}))].
\end{split}
\end{align}

Since we evaluate the robustness of the encoded features from a well-trained TSC transceiver, we have $p_\varphi(Z|X)$ and $D_\theta(\Tilde{Z})$. Besides, the corrupted features with artificial noise are given by $\Tilde{Z}\sim \mathcal{N}(E_\varphi(X),\sigma^2)$. Thus, we can obtain the upper bound of $I(\Tilde{Z},Y)$. 

Next, we focus on the second mutual information, $I(\Tilde{Z}, X)$, in (\ref{eq:opt2}) and have
\begin{align}
\begin{split}
I(\Tilde{Z},X)&=\int p(\Tilde{z},x)\log\frac{p(\Tilde{z},x)}{p(\Tilde{z})p(x)}d\Tilde{z}dx\\
&=\int p(\Tilde{z},x)\log \frac{p(\Tilde{z}|x)}{p(z)}dzd\Tilde{z}dx \\
&\ \ \ +\int p(\Tilde{z},x)\log \frac{p(z)}{p(\Tilde{z})}dzd\Tilde{z}dx\\
&=KL[p(\Tilde{Z}|X)||p(Z)]-KL[p(\Tilde{Z})||p(Z)],
\end{split}
\end{align}
where $KL$ represents Kullback–Leibler divergence that measures the difference between two probability distributions. Since $KL[p(\Tilde{Z})||p(Z)] \geq 0$, we have 
\begin{align}
\begin{split}
I(\Tilde{Z},X) & \leq KL[p(\Tilde{Z}|X)||p(Z)]\\
&=\frac{1}{2}\sum_{k=1}^m[\frac{\sigma_{k}^2}{\delta_k^2}+\log \frac{\delta_k^2}{\sigma_{k}^2}-1],
\end{split}
\end{align}
where $k$ denotes the index of the artificial noise variation added to the $k$th encoded feature unit, \textit{i.e.}, $\sigma=[\sigma_1, \cdots, \sigma_m]$, and $\delta=[\delta_1, \cdots, \delta_m]$ represents the inherent variation of an input reflected in the encoded feature space for task $\mathcal{T}$.

Built on the above derivation, we obtain the upper bound of the objective function in (\ref{eq:opt2}) as
\begin{align}\label{eq:loss}
\begin{split}
L(\sigma)\leq & \underset{X\sim p(X),Z\sim p_\varphi(Z|X), \Tilde{Z}\sim \mathcal{N}(E_\varphi(X),\sigma^2)}{\mathbb{E}}[-\mathcal{L}(Y,D_\theta(\Tilde{Z}))]
\\ & -\beta (\frac{1}{2}\sum_{k=1}^n[\frac{\sigma_{k}^2}{\delta_k^2}+\log \frac{\delta_k^2}{\sigma_{k}^2}-1]).
\end{split}
\end{align}

Therefore, by propagating $\Tilde{Z}$ through the decoding function, the artificial noise can be optimized by $\sigma=\sigma-\frac{ \partial {L}(\sigma) }{ \partial \sigma }$. 

\subsection{Robustness Mask and Case Study}
After optimizing artificial noise $\sigma$, we analyze the artificially corrupted encoded features $\Tilde{Z}$ and assess the robustness of each feature unit based on its sensitivity to semantic inference. Define the encoded feature variation of task $\mathcal{T}$ by $R = \max (\delta^2)$, which represents the maximal input variations mapping to the encoded feature space $Z$. Intuitively, the injected artificial noise should be restricted below $R$ to ensure inference reliability. However, recent research indicates that the correlation between different units in the feature space and inference performance are different~\cite{tsipras2018robustness}. Thus, feature unit $z_k\in z=\{z_1, \dots, z_m\}$ with a high correlation to inference results should have $\sigma_k^2>R$, \textit{i.e.}, robust against channel impairment, and  $z_k$ with $\sigma_k^2<R$ is non-robust since small channel impairment behaves as a strict restriction to retain semantic inference performance. Therefore, we explicitly disentangle encoded features into robust and non-robust against channel variations for TSC. 

\vspace{0.5em}
\textit{\textbf{Remark:}}
{By optimizing $\sigma_k\in\sigma$ for each unit of the encoded features, the IB-based problem formulation in (\ref{eq:opt2}) controls the information flow to the decoding function and evaluates feature unit robustness. For task $\mathcal{T}$, we approximate $X\sim p(X)$ with empirical risk minimization. Hence, the soft robustness mask of feature unit $z_k\in z = \{z_1,\dots, z_m\}$ is given by 
\begin{align}\label{eq:mask}
    r_k = \frac{\sum_{x_i\in\mathcal{X}_T}\sigma_k^i}{\sum_{x_j\in\mathcal{X}_T}\sum_{l=1}^m\sigma_l^j}, \ \ \ \sum_{k=1}^m r_k =1.
\end{align}

This mask can be leveraged by a well-trained TSC transceiver (\textit{i.e.}, given $E_\varphi$ and $D_\theta$) for a specific task $\mathcal{T}$ to accommodate instantaneous channel variations. Based on the robustness score, priority is provided for transmitting the encoded feature units to achieve reliable semantic inference. 
}
\vspace{0.3em}

We conduct a case study to leverage the robustness mask $r$ for subchannel allocation. 
Feature units with a small robustness score are considered non-robust against channel impairments, which should be assigned to high-quality subchannels that are measured by channel estimation techniques~\cite{coleri2002channel}. Given a total of $s$ available subchannels with distinct subchannel conditions, we leverage a greedy-based method to assign the $s$ subchannels to the $m$ encoded feature units based on their robustness mask $r$. The allocation problem is thus redefined as a direct one-to-one pairing problem, which is greedily solved with a computational complexity of $O(m)$, where $m$ represents the count of encoded feature units. The pseudocode of the robustness mask-based subchannel allocation is given in Algorithm~\ref{alg1}.

\begin{algorithm}[t!]
\caption{Channel-Resilient TSC: Subchannel Allocation} 
\label{alg1}
\begin{algorithmic}
\REQUIRE (1) a well-trained TSC transceiver with encoding function $E_\varphi$ and decoding function $D_\theta$; (2) input data and its corresponding semantic information $(x,y)\in(\mathcal{X}_T,\mathcal{Y}_T)$; (3) instantaneous $CSI = \{CSI_1,\dots,CSI_s\}$ for $s$ subchannels;  
\\
\ENSURE (1) robustness mask (2) subchannel assignment \\
// \textit{Robustness mask generation}\\
\begin{enumerate}
    \item \textbf{Encode}: $z=E_\varphi(x)=\{z_1,\dots,z_m\}$, $x\in\mathcal{X}_T$.
    \item \textbf{Initialize artificial noise}: $\sigma = 0$ with a neutral noise.
    \item \textbf{For each iteration}:
    \begin{enumerate}
        \item \textbf{Adjust noise}: $\sigma = \log(1 + \exp(\sigma))$, ensure positive through SoftPlus.
        \item \textbf{Inject noise}: $\Tilde{z}=z+\sigma \cdot \epsilon$ based on (\ref{eq:opt2}).
        \item \textbf{Decode}: $\hat{y} = D_\theta(\Tilde{z})$.
        \item \textbf{Update}: $\sigma=\sigma-\frac{ \partial {L}(\sigma) }{ \partial \sigma }$ based on (\ref{eq:loss}).
    \end{enumerate}
    \item \textbf{End For}
    \item \textbf{Robustness mask}: $r=\{r_1, \dots, r_m\}$ based on (\ref{eq:mask}).
\end{enumerate}
\textit{// Robustness mask implementation: assign subchannels}
\begin{enumerate}
        \item \textbf{For iteration $l\in \{1,\dots, m\}$ do}:
        \begin{enumerate}
        \item Select the smallest $r_i\in r$ and  remove $r_i$ from $r$;
        \item Select the best $CSI_j \in CSI$;
        \item Assign $z_i$ to the $j$th subchannel; 
        \item If the $j$th subchannel is fully occupied, remove it from $CSI$;
        \end{enumerate}   
        \item \textbf{End for}
\end{enumerate}
\end{algorithmic}
\end{algorithm}

\section{Numerical Evaluation}
\subsection{Evaluation Setup}
We consider two tasks for TSC on image classification, \textit{i.e.}, CIFAR-10~\cite{com11} and Street View House Numbers (SVHN)~\cite{com12} datasets. 
We adopt the VGG16~\cite{com13} architecture as the encoding function $E_\varphi$ and three fully connected layers as the decoding function $D_\theta$. Without loss of generality, we use the widely used additive white Gaussian noise (AWGN) channel as the general channel model to train the TSC transceiver, \textit{i.e.}, the encoding and decoding functions, $E_\varphi$ and $D_\theta$, that are frozen for channel resilience analysis. 
Besides, we consider a typical OFDM symbol structure with a total of 272 subcarriers. For each OFDM symbol, 16 pilots are evenly distributed across 256 data subcarriers~\cite{liu2024ofdm}. The number of encoded feature units is 512. The hyperparameter $\beta$ of IB is empirically set as $0.3$. 

\begin{figure}[t!]
 \vspace{-1em}
 \centering
	\subfloat[\centering First-half feature units, ideal channel]{\includegraphics[width=.36\columnwidth]{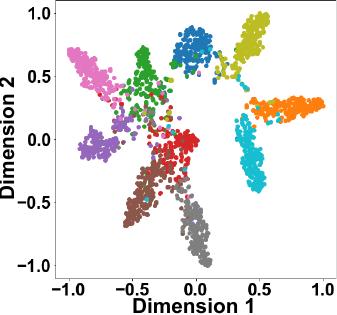}}\hspace{5pt}
	\subfloat[\centering Second-half feature units, ideal channel]{\includegraphics[width=.36\columnwidth]{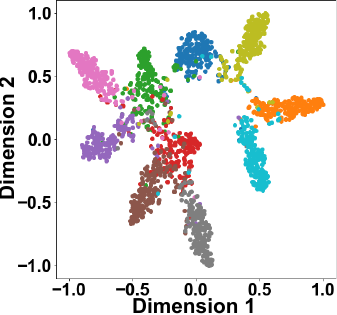}}\\
	\subfloat[\centering First-half feature units, noisy channel]
 {\includegraphics[width=.36\columnwidth]{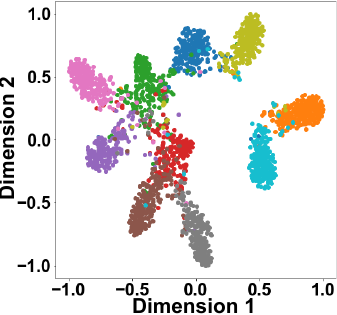}}\hspace{5pt}
	\subfloat[\centering Second-half feature units, noisy channel]{\includegraphics[width=.36\columnwidth]{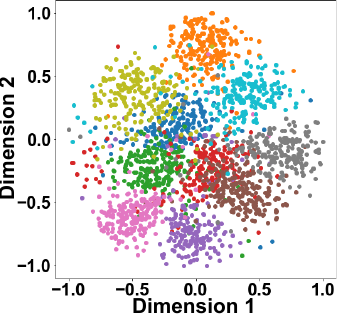}}
	\caption{Comparison of feature-level channel resilience between the ideal (noise = 0) and noisy (SNR = 0) channel conditions. 
 }\label{fig:cluster}
\end{figure}


\subsection{Evaluation Results}

\begin{figure}[t]
	\centering
	\subfloat[High Variance, CIFAR-10]{\includegraphics[width=.44\columnwidth]{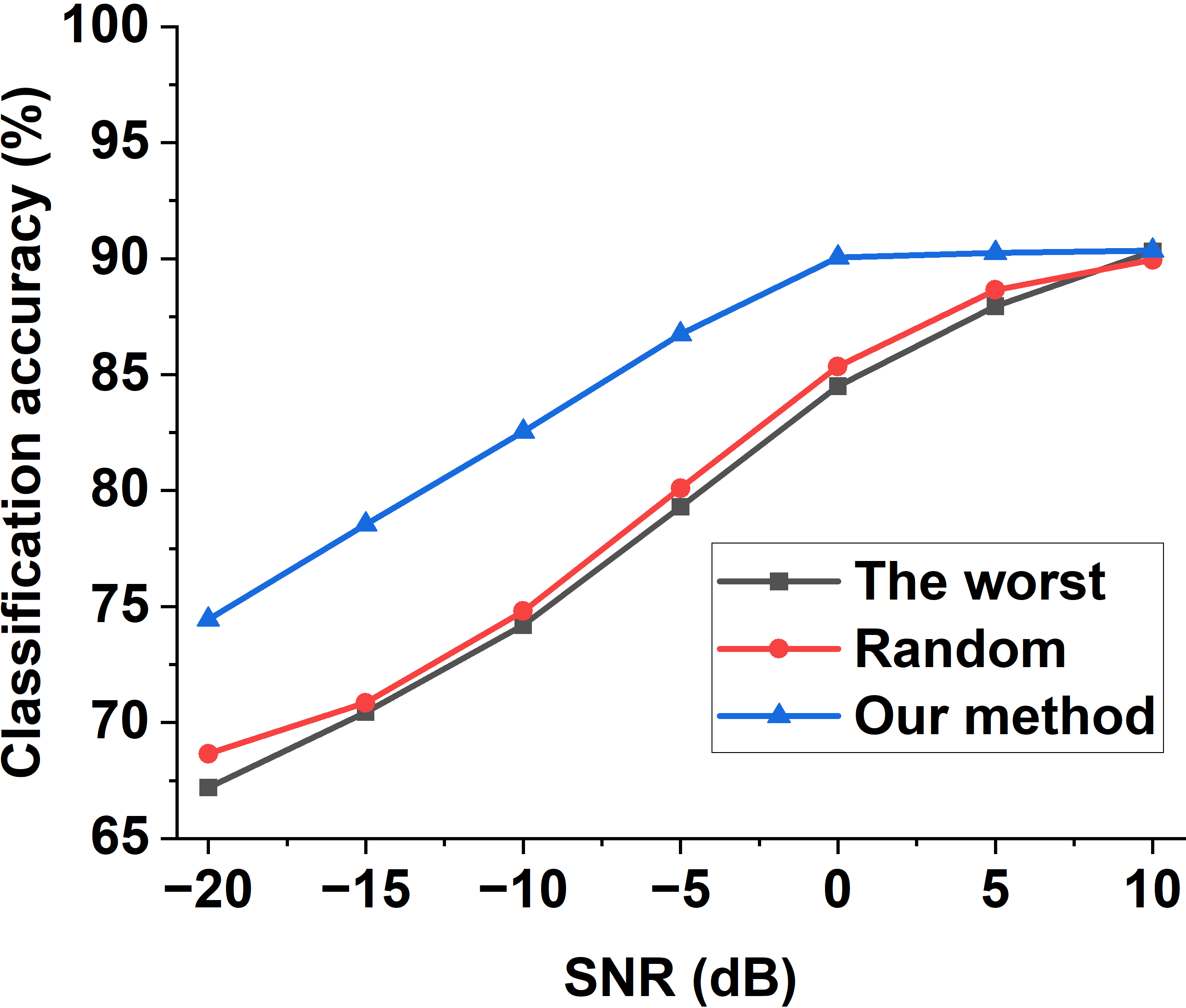}}\hspace{5pt}
	\subfloat[Low Variance, CIFAR-10]{\includegraphics[width=.44\columnwidth]{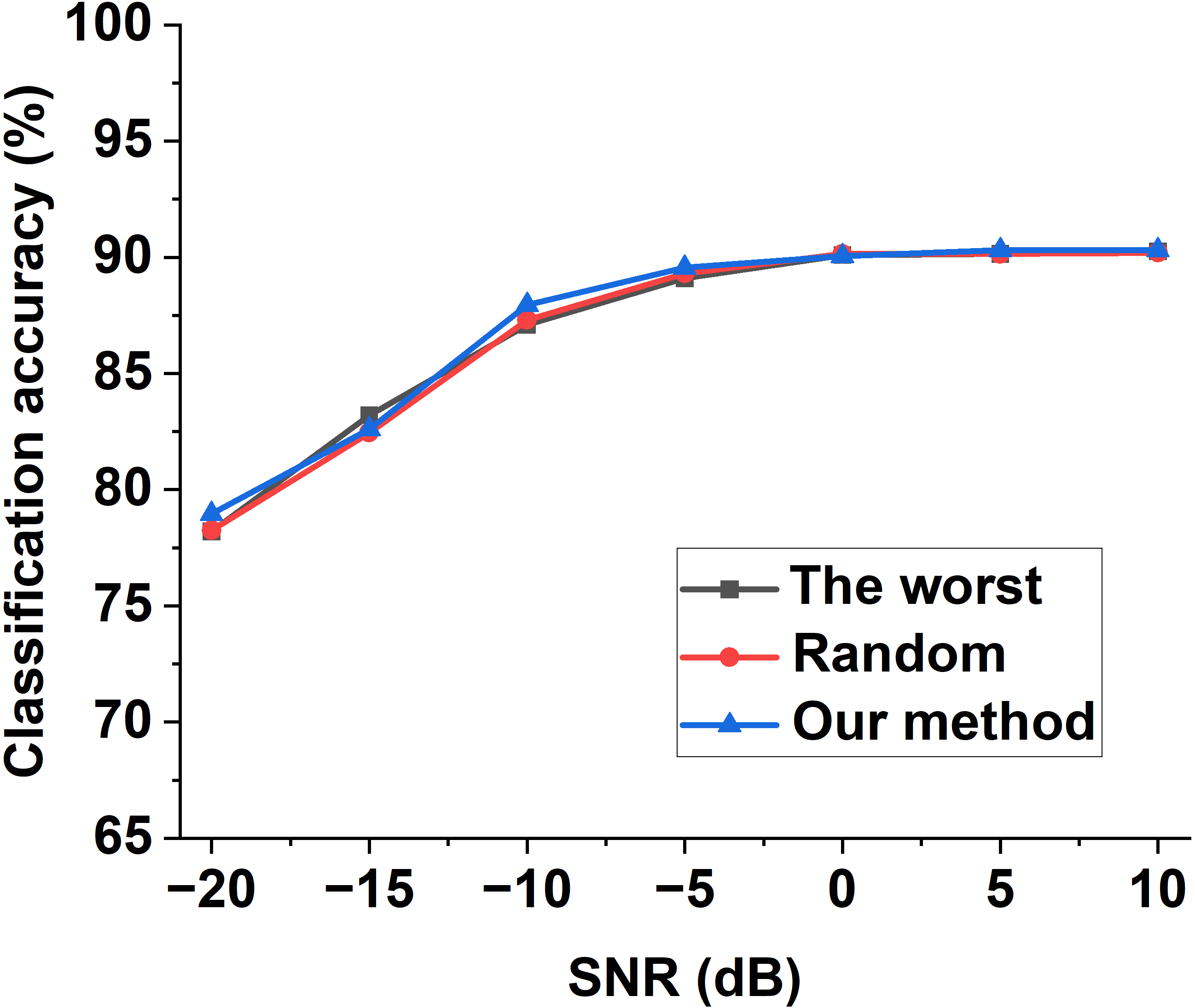}}\\
	\subfloat[High Variance, SVHN] 
 {\includegraphics[width=.44\columnwidth]{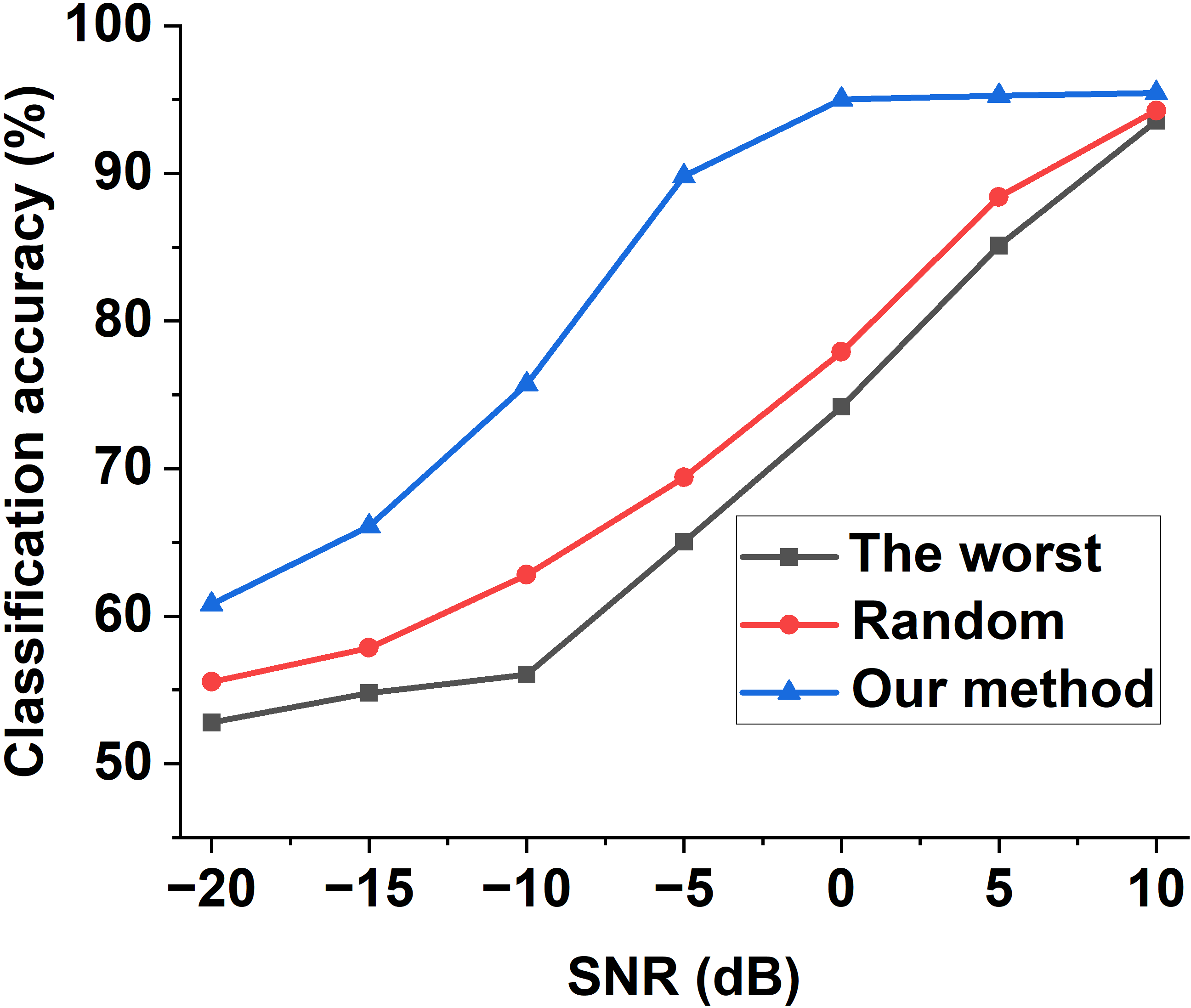}}\hspace{5pt}
	\subfloat[Low Variance, SVHN] 
 {\includegraphics[width=.44\columnwidth]{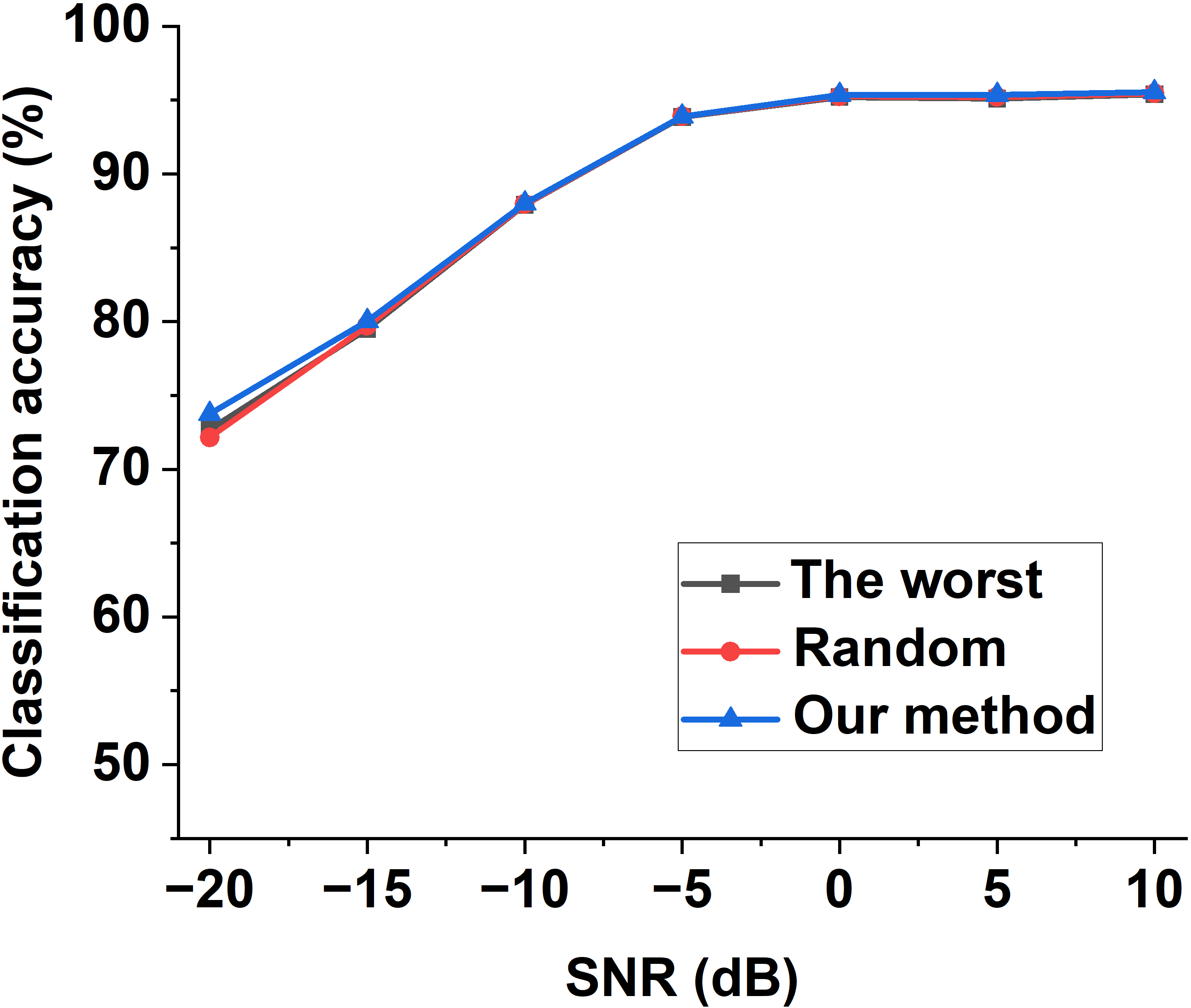}}
	\caption{Comparison of inference performance between dynamic (high variation, 15) and stable (low variation, 2) subchannel environments on CIFAR-10 and SVHN tasks. 
 }\label{fig:accuracy}
\vspace{-1em}
\end{figure}

We first evaluate the effectiveness of the robustness mask for encoded feature units on the CIFAR-10 task. According to the robustness scores, we rerank encoded feature units.  The units in the first half, which have higher scores, are expected to be more resilient against channel variations than those in the second half, which have lower scores. Fig.~\ref{fig:cluster} visualizes the feature-level inference performance between ideal and noisy channel conditions using 2D t-SNE. 
In ideal channel environments,\textit{ i.e.}, no noise, both robust (first-half) feature units and non-robust (second-half) feature units achieve comparable inference performance. However, under noisy conditions, \textit{i.e.}, SNR = 0, the performance disparity between the robust and non-robust feature units becomes apparent. Despite channel impairments, the first half of the units with higher robustness scores maintain performance levels similar to those in ideal conditions ((a) vs. (c)), whereas the performance of the second half deteriorates significantly ((b) vs. (d)). This contrast demonstrates the effectiveness of the robustness mask in evaluating feature-level channel resilience.

We then evaluate the effectiveness of using the robustness mask for subchannel allocation between two AWGN environments characterized by low and high variations. The low variation represents stable subchannel environments, while the high variation denotes highly dynamic discrepancies between subcarriers, regardless of their average performance. Fig.~\ref{fig:accuracy} illustrates the inference accuracy with on-average subchannel performance (SNR) between different channel variances on two image classification tasks. Specifically, Fig.~\ref{fig:accuracy}(a) and Fig.~\ref{fig:accuracy}(c) present inference performance under highly dynamic subchannel conditions. We compare our method against two baselines: random and worst-case allocations, the latter assigning the lowest-quality subchannels (low SNR) to the least robust feature units (low robustness score).
We observe that our method consistently outperforms the baselines across all SNR levels, which demonstrates its effectiveness. As the SNR increases, the performance gap between our method and the baselines narrows on both datasets. This suggests that satisfactory average subchannel performance can compensate for subchannel variations. 
\textit{The advantage of our method is particularly notable in highly dynamic and challenging channel conditions}. 
Furthermore, by comparing Fig.~\ref{fig:accuracy}(b) and Fig.~\ref{fig:accuracy}(d), we observe a marginal difference between the three methods in stable channel environments. 
This consistently validates our finding above. 

\section{Conclusion}
This letter introduced an innovative framework for channel-resilient task-oriented semantic communications (TSC), which analyzes the encoded feature space of a well-trained TSC transceiver. A robustness mask for encoded feature units is created based on the information bottleneck, which was implemented for subchannel allocation as a case study. The effectiveness of this framework was validated on two TSC tasks. To the best of our knowledge, this letter is the first to explore feature-level channel resilience for TSC against instantaneous channel variations. 

\bibliographystyle{IEEEtran}
\bibliography{main}

\end{document}